# Ambipolar Transport in Narrow Bandgap Semiconductor InSb Nanowires

B. Dalelkhan[1], D. Göransson[1], C. Thelander[1], K. Li[2], Y. J. Xing[2], V. F. Maisi[1], and H. Q. Xu[1,2,3,*]

[1]NanoLund and Division of Solid State Physics, Lund University, Box 118, S-22100 Lund, Sweden

[2]Beijing Key Laboratory of Quantum Devices, Key Laboratory for the Physics and Chemistry of Nanodevices, and Department of Electronics, Peking University, Beijing 100871, China

[3]Beijing Academy of Quantum Information Sciences, Beijing 100193, China

[*]Corresponding author. Email: hqxu@pku.edu.cn

(January 18, 2020)

We report on transport measurement study of top-gated field effect transistors made out of InSb nanowires grown by chemical vapor deposition. The transistors exhibit ambipolar transport characteristics revealed by three distinguished gate-voltage regions: In the middle region where the fermi level resides within the bandgap, the electrical resistance shows an exponential dependence on temperature and gate voltage. With either more positive or negative gate voltages, the devices enter the electron and hole transport regimes, revealed by a resistance decreasing linearly with decreasing temperature. From the transport measurement data of a 1-μm-long device made from a nanowire of 50 nm in diameter, we extract a bandgap energy of 190-220 meV. The off-state current of this device is found to be suppressed within the measurement noise at a temperature of $T$ = 4 K. A shorter, 260-nm-long device is found to exhibit a finite off-state current and a hole, on-state, circumference-normalized current of 11 μA/μm at $V_D$ = 50 mV which is the highest for such a device to our knowledge. The ambipolar transport characteristics make the InSb nanowires attractive for CMOS electronics, hybrid electron-hole quantum systems and hole based spin qubits.

Semiconductor InSb is an attractive material system for applications in high-speed, low-power electronics[1] and infrared optoelectronics[2] due to its high electron mobility, narrow bandgap and low electron effective mass. Moreover, its strong spin-orbit interaction and large g-factor make it one of the best-suited materials to create Majorana fermions in condensed matter systems.[3,4] In addition, gate tunable ambipolar transport measurements of InSb nanowire (NW) quantum dots show that InSb NWs are of emerging materials systems for achieving hole based spin qubits.[5] InSb NWs with a diameter smaller than 50 nm have shown to reach the quantum capacitance limit, leading to a significant improvement in device performance.[6] As the InSb NW diameter decreases from 65 nm to 10 nm, the bandgap increases from 0.17 eV to 0.4 eV, allowing tunable infrared detection in a range of 3.1 μm to 7.3 μm in wave length.[7] Previous studies show also that spin coherence length increases with decreasing the NW diameter, which has technological importance for spin-based devices.[8] Despite the fact that InSb NWs with a small diameter provide the above benefits, it has been challenging to grow InSb NWs with a diameter smaller than 60 nm by standard growth techniques, such as molecular beam epitaxy, metal-organic vapor-phase epitaxy[9-10] and chemical beam epitaxy. [11]

In contrast to the above growth techniques, chemical vapor deposition (CVD) offers possibilities for synthesis of NWs with tuned dimensions in wide ranges, including InSb NWs with diameters down to 10 nm and lengths up to tens of micrometers.[12-14] Single crystalline InSb NWs have been synthesized using CVD on InSb substrates[7,12-14] and initial electrical measurements of these NWs at room temperature show either the n-type[12-14] or the p-type[7] transport characteristics. However, electrical characterization of InSb NWs grown by CVD has so far been limited mainly to room temperature measurements and tuning of the carrier transport from n-type to p-type has been very difficult due to the use of a global back gate that does not couple sufficiently strong to the NWs. Thus, a systematic temperature-dependent transport study of CVD-grown InSb NWs under a sufficiently strong gate coupling has been lacking until now.

In this work, we use InSb NWs grown by CVD to fabricate top gate field effect transistors (FETs). In contrast to the previous works,[7,12-14] we obtain efficient gating by employing an Ω-shaped gate as shown in Figures 1(a) and 1(b). This combined with a small bandgap in InSb allows us to switch the NW FETs between the electron and the hole transport and, therefore, to achieve ambipolar transistor operation. We carry out temperature dependent transport measurements of the devices with a 1-μm-long and a 260-nm-long channel. From the measurements of the 1-μm-long channel device, we extract a bandgap energy of 220 meV in an InSb NW with a diameter of 50 nm. We also extract the field effect mobility of holes and electrons in the NW and study their temperature dependences. For the short, 260-nm-long InSb NW channel device, we demonstrate a very high, on-state hole current of 11 μA/μm (circumference-normalized) at a source-drain bias voltage of $V_D$ = 50 mV.

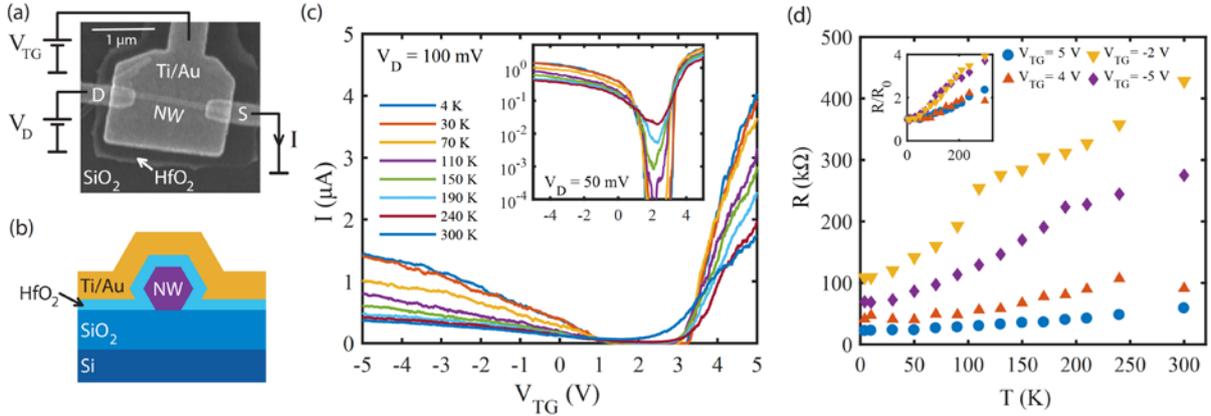

*Figure 1.* *(a) SEM image of the top-gated InSb NWFET measured in this work. A 10 nm thick $HfO_2$ is employed as a dielectric between the NW and the top gate metal electrode which is made of 7-nm-thick Ti and 128-nm-thick Au. The source and drain contacts consist of a 3-nm-thick layer of Ti and an 87-nm-thick layer of Au, and have a spacing of 1 μm. The NW has a diameter of 50 nm. The current I is measured from the source contact and the bias voltage $V_D$ is applied on the drain side as depicted in the circuit diagram. The top gate voltage $V_{TG}$ is swept to obtain the device transfer characteristics while the back gate is kept grounded in all the measurements. (b) Schematic view of the cross section of the top gated InSb NWFET. (c) Current I measured as a function of $V_{TG}$ at different temperatures at a source-drain bias voltage of $V_D$ =100 mV. The inset shows the same transfer characteristics but with current I obtained at $V_D$ =50 mV and plotted in logarithmic scale. (d) On-state resistance R at gate voltages selected from the electron and hole transport regions at $V_D$ = 100 mV. The inset shows the $R_0$-normalized resistance where $R_0$ is the low temperature resistance value at 4 K. Here it is seen that the normalized resistances measured in the electron transport region or in the hole transport region fall on top of each other, revealing that the temperature dependence of the resistance in each transport region does not depend on carrier concentration.*

The InSb NWs investigated in this work were grown by CVD on a Si substrate, see Ref. 15 for further details. The InSb NWFETs were fabricated on a degenerately n-doped Si substrate covered with a 200 nm thick layer of $SiO_2$ on top. A scanning electron microscope (SEM) image of a measured device is shown in Figure 1(a) with a schematic for its cross section given in Figure 1(b). The device has a channel length of 1 µm and a NW diameter of 50 nm. A 10-nm-thick layer of HfO$_2$ was used as the top gate dielectric material. In addition to the NW, the top gate partially covers the source and drain contacts to obtain efficient gating of the NW over the entire NW channel. The device fabrication started by mechanically transferring NWs from the growth Si substrate to the device Si/SiO$_2$ substrate. The substrate was cleaned by oxygen plasma etching before NW transferring, which normally leads to an improvement in the NW channel field effect mobility[16]. After determining the positions of NWs by SEM, we defined source-drain contacts using electron beam lithography (EBL). To achieve a low contact resistance, the NWs was etched in a diluted (NH$_4$)$_2$S$_X$ solution at 35°C in order to remove the native surface oxide on the NWs prior to the deposition of Ti/Au (3 nm/87 nm in thickness) metal electrodes via thermal evaporation. The gate dielectric was then fabricated by performing a second step of EBL and by growth of a 10-nm-thick layer of HfO$_2$ via atomic layer deposition. After lift-off, the top gate was fabricated by a third step of EBL, thermal evaporation of Ti/Au (5 nm/128 nm in thickness), and lift-off.

The fabricated devices were measured at temperatures of $T$ = 4.2 K to 300 K using the measurement circuit setup shown in Figure 1a. Here, a fixed DC bias voltage $V_D$ was applied to the drain contact, the source-drain current $I$ was recorded on the source contact side as a function of top gate voltage $V_{TG}$, and the back gate was grounded during all the measurements. Figure 1c depicts measured source-drain current $I$ of the device shown in Figure 1(a) as a function of $V_{TG}$ at a source-drain bias voltage of $V_D$ = 100 mV at different temperatures. The measurements show clear ambipolar transport characteristics with three distinct transport regions. In region I, 1.4 V ≤ $V_{TG}$ ≤ 3.2 V, the Fermi level E$_F$ resides within the bandgap of the channel material and the current $I$ is suppressed. In addition, the measurements at different temperatures show that the current in this region decreases exponentially with decreasing temperature, as seen in the inset of Figure 1(c). In region II with $V_{TG}$ < 1.4 V and region III with $V_{TG}$ > 3.4 V, the current $I$ increases with decreasing temperature and gate voltage span. In these two regions, the transistor enters the hole and electron transport regions, respectively. The maximum values of the circumference-normalized current $I_{on}/L$ obtained are 26 µA/µm in the electron transport region and 10 µA/µm in the hole transport region at 4 K. Here $L$ is the circumference of the NW. The on-state current $I_{on}$ in the electron transport region is higher than the values reported previously in Refs. 17-20, although our device has a smaller NW diameter and a larger contact spacing. The lowest current in region I, known as the off-state current, $I_{off}$ is also smaller than previously reported values[17,18] and decreased from 65 nA at 300 K to below noise level of 2-3 pA at 4 K. From the logarithmic plot in the inset of Figure 1(c), we extracted sub-threshold slope (SS) values of 75 mV/dec and 100 mV/dec in the electron and hole transport regions at 4K, respectively. The slightly larger value of SS in the hole transport region compared with the value in the electron transport region could be due to a lower mobility of holes and thus an overall reduced hole current. At room temperature, the SS values in the electron and hole transport regions are 454 mV/dec and 833 mV/dec, respectively. These are an order of magnitude larger than the ideal value of 60 mV/dec. The larger values most likely result from carrier excitations through the narrow bandgap of the material and/or from charge traps at the interface between the gate oxide material and the NW[18].

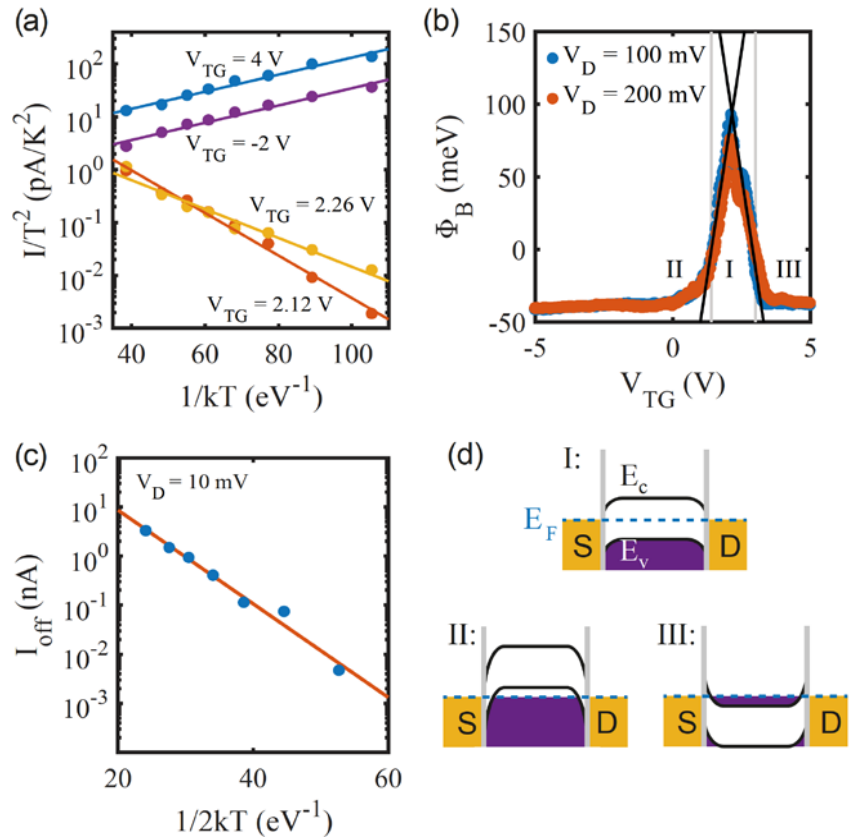

*Figure 2.* (a) Arrhenius plots for the current I vs temperature T of the same device as in Figure 1 at different values of $V_{TG}$ and at $V_D$ = 100 mV. (b) Effective energy barrier $\Phi_B$ extracted from the Arrhenius plots. (c) Off-state current $I_{off}$ in logarithmic scale as a function of temperature. From the slope of the line the extracted bandgap energy is 220 meV. (d) Energy band diagrams corresponding to the off-state region I, the hole transport region II, and the electron transport region III.

Figure 1(d) shows the temperature dependent measurements of the resistance $R$ of the device in regions II and III at $V_D$ = 100 mV. The resistance $R$ decreases with decreasing temperature $T$. A similar temperature dependence of the on-state resistance $R$ has been reported in both n-type InAs NWFETs [21] and n-type InSb NWFETs[22] and is explained by suppression of phonon scattering at decreased temperatures. It is seen that in both the hole transport region (region II) and the electron transport region (region III), the resistance showed a linear dependence on temperature, $R = R_o(1 + \alpha T)$, where the temperature coefficients $\alpha$ = 0.0068/K in the electron transport region and $\alpha$ = 0.0167/K in the hole transport region. Here $R_0$ is the low temperature resistance that we determined at $T$ = 4 K. As presented in the inset, $\alpha$ does not depend on carrier concentration, obtained at different $V_{TG}$ values. But, interestingly, it is different in the electron and hole transport regions. The linear temperature dependence of the resistance is typical for metallic or metallic-like systems, which proves that the NWFET is tuned deep into the valence band or the conduction band.

To extract the energy barrier $\Phi_B$ of the injection carriers from the source contact and the bandgap energy in the InSb NW, we analyze the current obtained at different temperatures at a given bias voltage $V_D$ using the equation of thermionic emission,[23]

$$I = AA^*T^2 \exp\left(\frac{-\Phi_B}{kT}\right), \quad (1)$$

where $A$ is the effective contact area and $A^*$ is the Richardson constant. Figure 2(a) shows representative curves of $I/T^2$ in logarithmic scale as a function of $1/kT$ extracted from the measurement data shown in Figure 1(c) at $V_{TG}$=4 V (region III, electron transport region), $V_{TG}$=-2 V (region II, hole transport region), and $V_{TG}$=2.26 and 2.12 V (region I, for the Fermi level in the bandgap). The solid lines in Figure 2(a) are the fits to the experimental data based on Eq. (1). Here, it is seen that straight lines are obtained in the fits and the slopes of these lines would give the effective energy barriers $\Phi_B$. We can extract the effective energy barriers $\Phi_B$ in the same way at all the gate voltages $V_{TG}$ considered. Figure 2(b) shows the extracted $\Phi_B$ as a function of $V_{TG}$ for $V_D$ = 100 mV and $V_D$= 200 mV. It is shown that the energy barrier reaches its maximum at $V_{TG}$ = 2.2 V. At this gate voltage, the Fermi level $E_F$ in the wire resides in the middle of the bandgap as shown in the energy band diagram I of Figure 2(d). Hence, we determine the bandgap as $E_g = 2\Phi_{B,MAX}$ = ~190 meV. We can also extract the energy bandgap in the NW from the activation energy of the carriers determined from the temperature dependent measurements of off-state current $I_{off}$ at a small source-drain bias voltage according to $I_{off} \sim e^{-\frac{E_g}{2kT}}$. Figure 2(c) shows the plot of the off-state current $I_{off}$ (in logarithmic scale) measured as a function of $1/2kT$ at $V_D$=10 mV. Here, a good straight line fit to the data is obtained, see the solid line in Figure 2(c). From the slope of the straight line, a bandgap of ~ 220 meV is extracted. The extracted bandgap energy of 190-220 meV is close to, but slightly higher than the bulk value of 170-180 meV, due to the quantum confinement effect. The slightly larger value of the bandgap determined by the thermal emission analysis when compared with the value determined by the carrier activation analysis is likely due to the fact that large bias voltages were applied in the measurements employed for the thermionic emission analysis. This is because a larger bias voltage should in general lead to a smaller extracted energy bandgap in the thermionic emission analysis.

As the gate voltage is tuned towards more positive or negative values, the effective energy barrier $\Phi_B$ of carrier injection from the contact is decreased, as the Fermi level in the NW moves closer to either the conduction or valence band. Finally, the Fermi level in the NW moves into the conduction or the valence band, as shown in the energy band diagrams II and III, respectively, in Figure 2(d), and the effective energy barrier vanishes. Deeper in the conduction and the valence band, we obtain negative values for the energy barrier, which indicates that the thermionic emission mechanism does not apply in these cases and direct tunneling of carriers contributes dominantly to the current. Based on the linear dependence of $\Phi_B$ on $V_{TG}$ in the region when the Fermi level resides in the bandgap, i.e., the black dashed lines in Figure. 2(b), we extract a gate level arm of $\alpha = 0.13 \frac{eV}{V}$ for holes and $\alpha = 0.097 \frac{eV}{V}$ for electrons. We observe a shoulder on the right side of the peak in Region I in Figure 2(b). A similar behavior was observed in Ref. 24 and was explained by charge trap states located within the bandgap. The origin of the charge trap states in the bandgap could be due to imperfect NW-oxide interfaces as studied in Ref. 25.

In the above analysis, we have neglected effect of the Schottky barriers on the carrier transport. Zhao et al.[26] analyzed the impact of Schottky barriers on narrow bandgap NWFETs. According to their analysis, the potential profile at a Schottky barrier can be described as $\phi_s(x) \sim e^{-x/\lambda}$ and the tunneling transparency of the Schottky barrier can be characterized by a barrier width $\lambda$ extracted at the fully depleted NW condition. For our nearly gate-around device, according to Ref. 27, $\lambda$ is given by

$$\lambda = \sqrt{\frac{\varepsilon_{NW} d_{NW}^2 \ln(1+\frac{2d_{ox}}{d_{NW}})}{8\varepsilon_{ox}}}, \quad (2)$$

where $\varepsilon_{ox}$ and $d_{ox}$ are the dielectric constant and the thickness of the gate dielectric HfO$_2$, and $\varepsilon_{NW}$ and $d_{NW}$ are the dielectric constant and the diameter of the InSb NW. Taking $\varepsilon_{ox}$ = 16-25, $d_{ox}$=10 nm, $\varepsilon_{NW} = 16.8$, and $d_{NW}$=50 nm, we obtain $\lambda$ = 8-10 nm, which indicates the barrier is very transparent based on the analysis in Ref. 26. Hence our assumption of neglecting the effects of the Schottky barriers is justified.

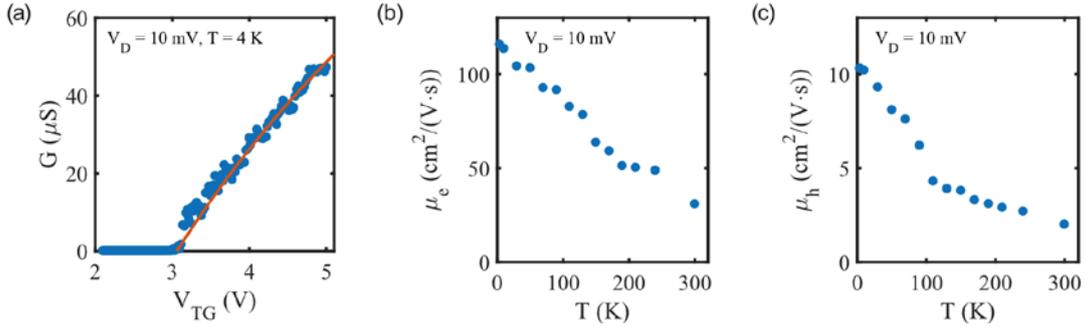

*Figure 3. (a) Conductance G measured for the same device as in Figure 1 as a function of top gate voltage $V_{TG}$ at $V_D$= 10 mV and T = 4 K. A field-effect mobility is extracted from a fit to the data (orange solid line) at $V_{TG}$ above $V_{th}$ using equation (3). (b) Temperature dependence of the electron field-effect mobility $\mu_e$. (c) Temperature dependence of the hole field-effect mobility $\mu_h$.*

We now evaluate the electron and the hole field effect mobility, $\mu_e$ and $\mu_h$, in our InSb NWFET. We extract the mobility $\mu$ from the measured conductance using the field effect mobility equation,

$$G(V_g) = \left(R_S + \frac{L^2}{\mu C(V_g - V_{th})}\right)^{-1}, \quad (3)$$

where $G$ is the conductance, $R_S$ the series resistance of the contacts, $L$ the channel length, $V_{th}$ the gate threshold voltage, and $C$ the gate capacitance.[16] The parameters $R_S$, $\mu$ and $V_{th}$ are the fitting parameters, whereas the capacitance $C$ is estimated from a cylindrical gate capacitor model,

$$C = \frac{2\pi\varepsilon\varepsilon_0 L}{\ln\left(\frac{h}{r}\right)}, \quad (4)$$

where $\varepsilon_0$ is the vacuum dielectric constant, $\varepsilon$ the relative dielectric constant of HfO$_2$, $r$ the radius of the NW, and $h$ the distance from the center of the NW to the gate electrode. Based on this equation, we obtain $C$ = 2.64 fF. A data fit of the measured electron conductance to Eq. (3) is shown in Figure 3(a). The blue points in the figure are the measured conductance values at temperature $T$=4 K and the solid orange line is the fit. Note that in the fit, the threshold voltage $V_{th}$ was first determined by the linear extrapolation from the currents measured at large positive gate voltages. The fit which is made for gate voltage $V_{TG}$ in a range of $V_{TG} \geq V_{th}$ yields $\mu$=~116 cm$^2$/Vs, $V_{th}$ = 2.9 V and $R_s$ = 2.1 kΩ at $T$= 4 K.

Figures 3(b) and 3(c) show the extracted electron and hole mobilities in the InSb NWFET at different temperatures. It is seen that the electron mobility increases from a value of $\mu_e$ = 31 cm$^2$/Vs at room temperature to a value of 116 cm$^2$/Vs at low temperatures. The hole mobility increases from $\mu_h$ = 2 cm$^2$/Vs at room temperature to 10 cm$^2$/Vs at low temperatures. The room temperature electron mobility extracted in this work is similar to the previously reported values for InSb NWs grown by CVD in Ref. 28. The Increase in the hole or electron mobility with decreasing temperatures is due to suppression of phonon scattering. Note that the relative change in the device resistance with temperature shown in Figure 1(d) is consistent with the extracted electron mobility change shown in Figures 3(b) and with the hole mobility change shown in Figure 3(c). However, both the electron and hole mobilities are lower than their corresponding bulk values. A lower carrier mobility has commonly been observed in a semiconductor NWFET and could be attributed to enhanced scattering by charge traps in the gate oxide and by imperfection at the NW-gate oxide interface.

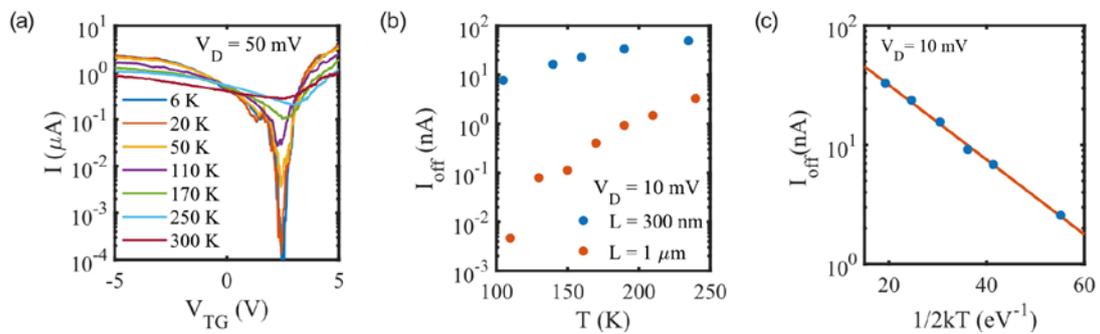

*Figure 4. (a) Transfer characteristics measured for a short channel InSb NWFET device with a channel length of L = 260 nm at a source-drain voltage of $V_D$ = 50 mV. (b) Off-state currents I$_{off}$ measured for the short channel device and for the long channel device as shown in Figure 1 as a function of temperature, respectively. (c) Off-sate current I$_{off}$ (in logarithmic scale) of the short channel device as a function of the temperature. The slope gives an effective bandgap energy of E$_g$ =145 meV.*

Figure 4 shows the measured transport properties of the short channel NWFET made from an InSb NW of $d$=65 nm in diameter with a channel length of $l$ = 260 nm. Figure 4(a) displays the transfer characteristics of the devices measured at $V_D$ = 50 mV and at various temperatures. This device again shows ambipolar transport characteristics. However, in contrast to the device discussed above, the on-state currents in the hole and electron transport regions are of similar magnitude. In addition, the currents are more than a factor of two higher than that in the device discussed above. In particular, we obtained a normalized on-state hole current of 11 µA/µm at $V_D$ = 50 mV. To our knowledge, this on-state hole current is the highest value ever reported for InSb NWFETs.[7,17,19] Figure 4b shows the measured off-state current of the short channel device measured at $V_D$=10 mV in comparison to the off-state current of the long channel device. It is seen that the off-state current of the short channel device is about three orders of magnitude higher than the off-state current of the long channel device. A high excess off-state current is often observed in a short channel, narrow bandgap semiconductor NWFET due to non-uniform or inefficient gating across the NW cross-section.[29] Such non-uniform gating in a top-gate NWFET could lead to band inclining along the cross section of the NW and thus the carriers see an effectively smaller bandgap in the NW channel. To demonstrate this, we in Figure 4(c) plot the off-state current (in logarithmic scale) of the short channel InSb NWFET as a function of *1/2kT* and to extract the effect bandgap in the NW channel. It is seen that the plot can be excellently fitted by a straight line and the slope of the line gives a bandgap of 145 meV, which is smaller than the bulk value of InSb, in consistence with our above analysis. This band inclining effect is less significant in our long channel NWFET which was made from a NW with a smaller diameter.

In summary, we have fabricated top-gate NWFETs from narrow bandgap semiconductor InSb NWs and demonstrated their ambipolar characteristics. The NWs are grown by CVD, and the top-gate devices are fabricated on a Si/SiO$_2$ substrate and are characterized by temperature dependent transport measurements. Thanks to having the top gate in close vicinity of the NW channel and the narrow bandgap of the channel material, the NWFETs can be efficiently tuned from their n-type on-states to their p-type on-states through their off-states and thus exhibit three well-distinguished transport regions in their ambipolar characteristics. The measurements show that the devices exhibit an n-type circumference-normalized on-state current of ~26 µA/µm, which is among the highest reported for InSb

NWFETs, and a p-type on-state current of ~11 µA/µm, which is the highest one ever reported so far for InSb NWFETs. From the temperature dependent measurements of the off-state current of the NWFET with a channel length of 1 µm and a NW diameter of 50 nm, a bandgap of 190-220 meV in the channel material is extracted. However, the corresponding measurements of the NWFET with a channel length of 260 nm and a larger NW diameter of 65 nm give an effective bandgap of 145 meV in the NW channel, which is smaller than the bulk value of InSb. The reason for extracting a smaller bandgap value in the NW channel is believe to be caused by non-uniform gating along the cross section of the NW. Our results presented in this work will provide a promising route towards developments of InSb NW based CMOS technologies and of a hole spin based platform for quantum information processing and computing.[5,30]

This work was supported by NanoLund, the Swedish Research Council (VR), the Ministry of Science and Technology of China through the National Key Research and Development Program of China (Nos. 2017YFA0303304 and 2016YFA0300601), the National Natural Science Foundation of China (Nos. 11874071, 91221202, and 91421303), and the Beijing Academy of Quantum Information Sciences (No. Y18G22).